# Effect of Temperature and Glycerol on The Hydrogen-Bond Dynamics of Water

|   |   |
|---|---|
| Authors | Pavan K GhattyVenkataKrishna, Edward C Uberbacher |
| Publication date | 2013/3/1 |
| Journal | CryoLetters |
| Volume | 34 |
| Issue | 2 |
| Pages | 166-173 |
| Publisher | Cryoletters |
| Description | The effect of glycerol, water and glycerol-water binary mixtures on the structure and dynamics of biomolecules has been well studied. However, a lot remains to be learned about the effect of varying glycerol concentration and temperature on the dynamics of water. We have studied the effect of concentration and temperature on the hydrogen bonded network formed by water molecules. A strong correlation between the relaxation time of the network and average number of hydrogen bonds per water molecules was found. The ... |
| Scholar articles | Effect of Temperature and Glycerol on The Hydrogen-Bond Dynamics of Water<br>PK GhattyVenkataKrishna, EC Uberbacher - CryoLetters, 2013<br>Related articles - All 3 versions |

# EFFECT OF TEMPERATURE AND GLYCEROL ON THE HYDROGEN-BOND DYNAMICS OF WATER

Pavan K. GhattyVenkataKrishna, Edward C. Uberbacher

Computational Biology and Bioinformatics Group, Oak Ridge National Laboratory, Oak Ridge, TN 37830, Email: pkc@ornl.gov

**Abstract**

The effect of glycerol, water and glycerol-water binary mixtures on the structure and dynamics of biomolecules has been well studied. However, a lot remains to be learned about the effect of varying glycerol concentration and temperature on the dynamics of water. We have studied the effect of concentration and temperature on the hydrogen bonded network formed by water molecules. A strong correlation between the relaxation time of the network and average number of hydrogen bonds per water molecules was found. The radial distribution function of water oxygen and hydrogen atoms clarifies the effect of concentration on the structure and clustering of water.

## INTRODUCTION

Water is essential for biological function and serves as the backdrop on which all the cellular processes take place. But it is seldom the lone solvent in biological systems. Along with water, a variety of co-solvents are present in varying concentrations. Various groups of researchers have studied the effect of co-solvents like glycerol [1-4], acetone [5], dimethyl-sulphoxide [6], acetonitrile [5, 7], methanol [8], urea [9] and formamide [10] on the structure and dynamics of water.

Arguably the most important property of water is its ability to form strong hydrogen bonds [11, 12]. The presence of co-solvents changes the behavior of water, specifically the hydrogen bonding pattern, spatial distribution and dynamics. It is known from literature [6, 9, 13, 14] that in the presence of co-solvents less polar than water, an increase in the concentration of the co-solvent leads to an increase in the clustering of water. Changes in the clustering pattern invariably affects the hydrogen-bond network of water. To elucidate these effects we have chosen glycerol as a co-solvent. Glycerol is an interesting candidate for a co-solvent since it has been shown to play a pivotal role in the freeze-tolerance of some species of frogs [15-18] and is considered to be one of the best cryo-protectant solvents[19].

Glycerol water binary mixtures have been studied using experimental techniques aimed at understanding their dielectric properties [1-4, 20]. Dielectric relaxation is a measure of the hydrogen bond rearrangement dynamics [11]. We know from these studies that at any given temperature (in the range 203-303K), as the glycerol content increases, the dielectric constant

(ε) decreases. Also, at any given concentration, as the temperature decreases, the ε increases. In another study, spin probe electron spin resonance was used to study vitrification of glycerol-water mixtures in the 21-97% wt glycerol concentration range and a sudden change in the free-volume of the mixture was observed at 77.3% wt glycerol which the authors interpreted as an inhomogeneous to homogenous transition in water hydrogen-bond network [21]. Molecular dynamics simulations and infrared spectroscopy have been employed to study glycerol-water binary mixtures [20, 22] and they conclude that high glycerol concentrations mimic the hydrogen bonding seen in ice without actually forming crystals. More recently, simulations of 42.9 and 60 wt % Glycerol in water at 25°C concluded that glycerol slowed the dynamics of water[23]. Furthermore, simulations such as the ones used in the current study were used to predict the glass transition temperature ($T_g$) of a 60% wt glycerol in water and excellent agreement with results from differential scanning calorimetry was found[24]. Simulations were also used to study the differences between glycerol-water (8-32% wt) and ethylene glycol-water (6-24% wt) binary mixtures in the 273-298K temperature range and concentration of solute was found to have a greater affect on hydrogen-bonding pattern than temperature[25].

## MATERIALS AND METHODS

We performed 66 molecular dynamics simulations of glycerol-water binary mixtures. 11 glycerol/water solutions were made with concentration by weight of glycerol ranging from 0% to 100% with increments of 10%. For each composition, six temperatures between 250K to 300K were simulated. The General AMBER Force Field [26] was used to model glycerol along with the TIP3P [27] water model. The simulation for each system consisted of 350ps of equilibration and 100ps of production data under a constant pressure of 1 atm. This is a reasonable choice of simulation length since 100ps is well within the time scale of the processes we intend to study.

## RESULTS AND DISCUSSION

The water-water hydrogen bonds in the system were analyzed using a hydrogen bond correlation function as described in Luzar and Chandler (1996).

$$c(t) = [h(0)h(t)]/[h(0)] \quad (1)$$

where, h(t) is a hydrogen bond operator which is 0 when the hydrogen bond is absent and 1 when present. The angled brackets indicate ensemble averages. Thus c(t) is the probability that a hydrogen bond formed at time t = 0 exists at time t = t. The structural relaxation time of this hydrogen bonded network $\tau_R$ is the time at which c(t) = 1/e. A hydrogen bond is supposed to be formed when for a pair of potential donor and acceptor, the distance between O–O is ≤ 3.4 Å and the O–H· · ·O angle is ≥ 150°. Figure 1 illustrates the effect of concentration and temperature on $\tau_R$ which is the log scale. Dark red and dark blue indicate high and low values of $\tau_R$ respectively. $\tau_R$ increases with an increase in the concentration of glycerol and/or a decrease in temperature. Putting it differently, the dynamics of water can be slowed down by decreasing the temperature, introducing glycerol, or a combination of both.

It is known from experiments that $\tau_R$ follows power law in the temperature range examined in the current study [28, 29]. To validate our results, we fitted the data with $T_c$ = 197.5K for comparison with findings from mode-coupling theory (MCT) and with $T_o$ = 160K for comparison with Vogel-Fulcher-Tammann (VFT) forms and the result (Figure 2) was an almost–linear fit for both theories. Similar results were obtained for the case of pure water[30].

We see from Figure 1 that the dependence on concentration of glycerol is more pronounced at lower temperatures. This can be explained by the increase in hydrogen bond strength in a more hydrophobic environment. Since glycerol is less polar than water, the polar distractions to a given water-water hydrogen bond are reduced in the presence of increasing glycerol content.

Given a significant change of water dynamics, it is expected that there would be a concomitant change of water structure. We can quantify this by calculating $N_H$, the average number of hydrogen bonds a water molecule makes with itself at a given concentration of glycerol w and temperature TK. $N_H$ was calculated by averaging the total number of hydrogen bonds water is involved in during the entire simulation with the total number of water molecules and the number of time steps. $N_H$ of 3.6 at 300K for pure water is in excellent agreement with literature [27]. Evidently, $N_H$ decreases with increase in concentration of glycerol and/or increase in temperature. The plot $\tau_R$ vs. $N_H$ can be fit using points at constant concentration (Figure 3) or constant temperature. The dotted lines on Figure 3 were obtained by fitting the data with a function of the type

$$\ln(\tau_R) = m*N_H + p \qquad (2)$$

where m and p are fit parameters; m appears to be a function of concentration.

To identify the energetic basis behind the dependence of $\tau_R$ on glycerol concentration, we calculated the water-water hydrogen bond energy was calculated using [31].

$$E = 25.3 \times 10^3 \exp[-3.6\, d(H \cdots O)] \text{ KJ/mol} \qquad (3)$$

where $d(H \cdots O)$ is the distance between the donor hydrogen and the acceptor oxygen. As can be seen from Figure 4, changes in temperature and concentration of glycerol contribute to a relatively small change in the energy of a hydrogen bond made by water. An 8% increase is seen from no glycerol to 90% glycerol. This allows us to conclude that the changes to the hydrogen bond network of water brought on by temperature and glycerol are primarily of entropic origin.

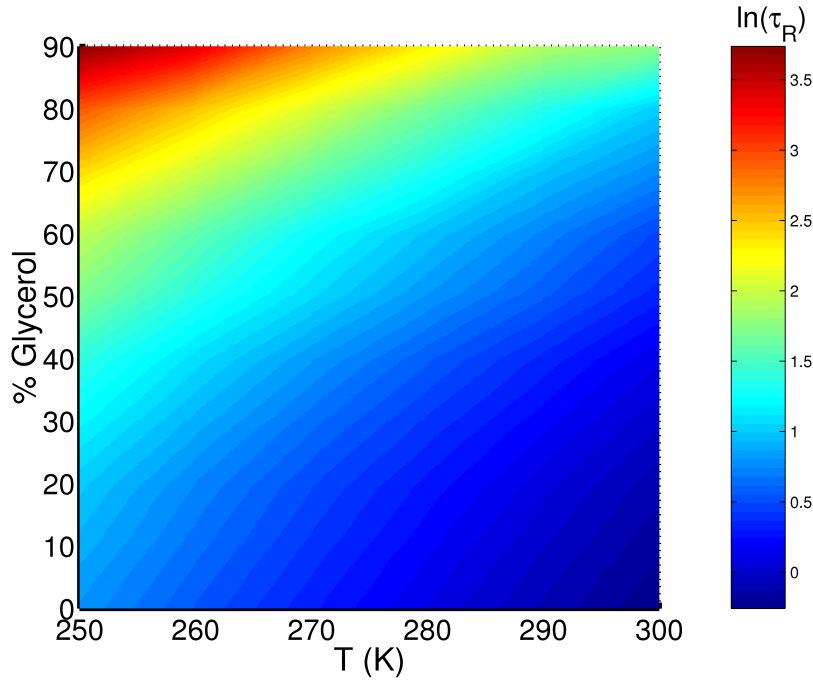

**Fig. 1**. Variation of $\tau_R$(ps) as a function of temperature (in K) and concentration (in % wt glycerol in water)

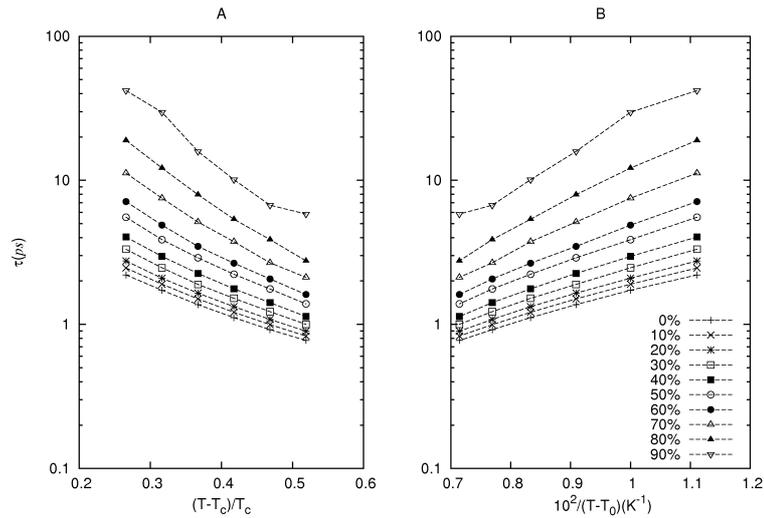

**Fig. 2.** Relaxation time $\tau_R$(ps) (A) Fit to scaling predicted by MCT with $T_c = 197.5K$ (B) Fit to VFT form with $T_0 = 160K$

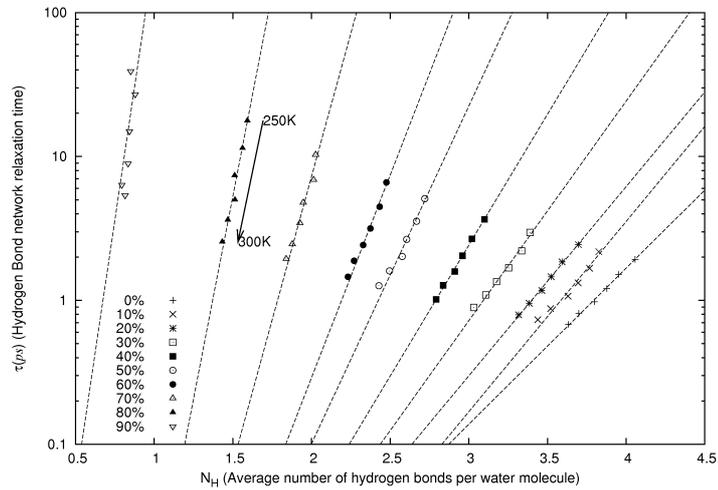

**Fig. 3.** Semi-log plot of $\tau_R$ vs. $N_H$ of water-water hydrogen bonds.

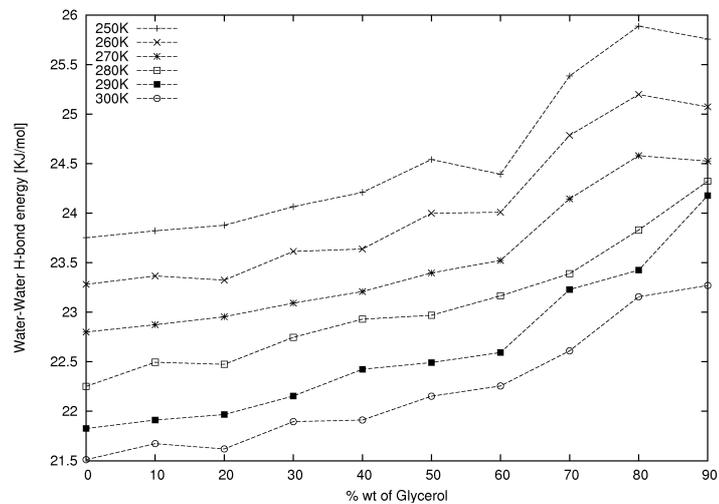

**Fig. 4.** Water-Water hydrogen bond energy in KJ/mol. Results here are similar to Stumpe and Grubmüller (2007).

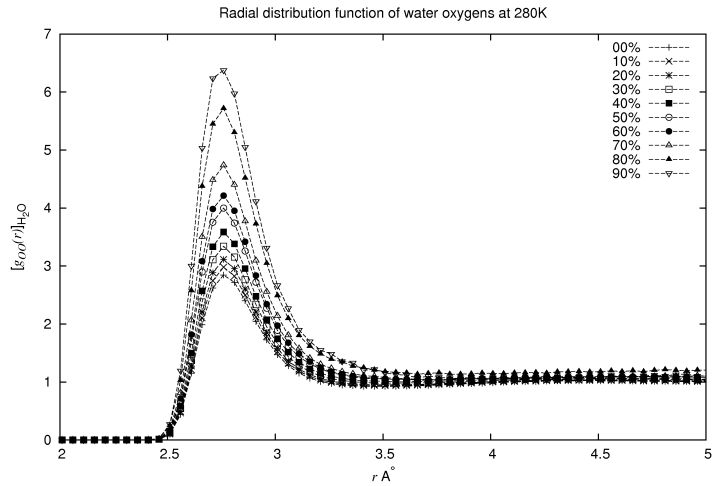

**Fig. 5.** Radial distribution function of water oxygens at 280K

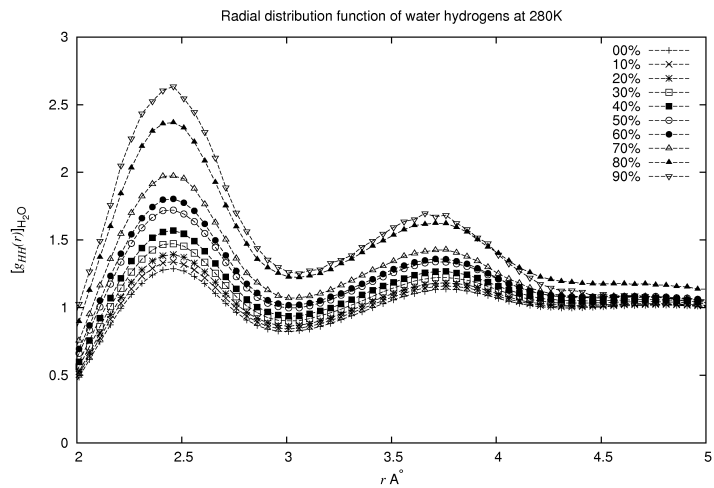

**Fig. 6.** Radial distribution function of water hydrogens at 280K

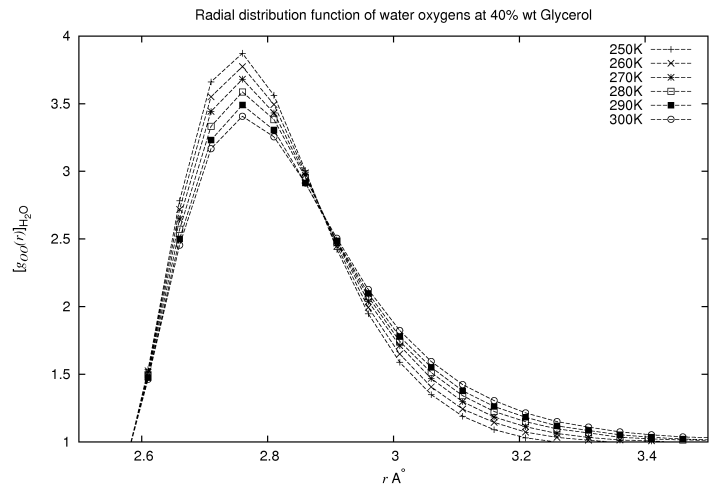

**Fig. 7.** Radial distribution function of water oxygens at 40% wt glycerol

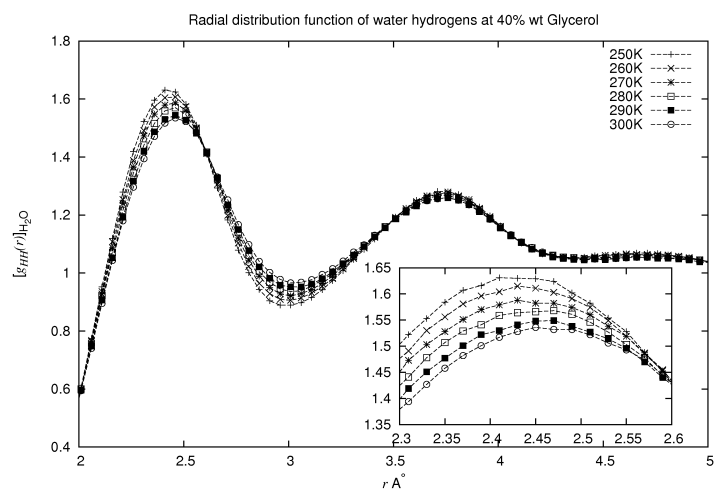

**Fig. 8.** Radial distribution function of water hydrogens at 40% wt glycerol. Inset shows the region between 2.3 − 2.6Å.

We also calculated the radial distribution function of water oxygens gOO(r) and water hydrogens gHH(r) and the data at 280K is plotted in Figures 5 and 6 as an example. Although the position of the first peak of gOO(r) at ~ 2.7Å is maintained at all the concentrations, the height of the peak increases with glycerol content which is indicative of water clustering. Formation of these clusters in turn means that the hydrogen bond network formed by the O-O pair relaxes at longer times which is precisely what is observed (Fig. 1). gHH (r) has three peaks at 1.5, 2.4, and 3.7Å where the latter two indicate a tetrahedral structure of water. gHH(r) follows a trend similar to gOO(r) (Figure 6). Temperature, on the other hand, seems to have very little effect on the radial distribution functions. Data for gOO(r) and gHH(r) at 40% glycerol has been plotted in Figures 7 and 8. A small decrease in the peak height with increase in temperature is noted in both cases. This supports the argument in the previous sections that the changes to hydrogen bond network of water are of entropic origin.

In conclusion, molecular dynamics simulations have been carried out in a systematic way for a series of glycerol-water solutions. These calculations confirm that the structure and dynamics change significantly with addition of glycerol. These changes are primarily of entropic origin. The water-water hydrogen bond becomes more persistent at higher concentration of glycerol. This is largely due to the decrease in the number of available water-water hydrogen bonds. Another important finding is that the dependence on concentration is more pronounced at lower temperatures and the dependence on temperature is more pronounced at higher concentrations. These findings have important ramifications in understanding the thermodynamic and dynamic behavior of proteins in aqueous solution. More rigidified water structure can damp the motions, thus helping stabilize the protein in solution and pave way for long-term preservation.

**Acknowledgements:** The research was sponsored by the US DOE under Contract No. DE-AC05-00OR22725 with UT-Battelle LLC managing contractor for the Oak Ridge National Laboratory.